\documentclass[tightenlines,aps,
twocolumn,
showpacs,nofootinbib]{revtex4}

\usepackage{psfig,float}
\usepackage{amsmath,amsfonts,graphicx,bm}

\newcommand{\ep}{\epsilon}
\newcommand{\be}{\begin{equation}}
\newcommand{\ee}{\end{equation}}
\newcommand{\ba}{\begin{eqnarray}}
\newcommand{\ea}{\end{eqnarray}}

\begin{document}

\begin{titlepage}

\begin{flushright}
\vbox{
\begin{tabular}{l}
 SLAC-PUB-10673\\
 UH-511-1056-04\\
 hep-ph/0409088
\end{tabular}
}
\end{flushright}

\title{
Higgs boson production at hadron colliders:
differential cross sections through
next-to-next-to-leading order
}

\author{
Charalampos Anastasiou\thanks{e-mail:babis@slac.stanford.edu}} 
\affiliation{
          Stanford Linear Accelerator Center,\\ 
          Stanford University, Stanford, CA 94309}
\author{Kirill Melnikov
        \thanks{e-mail: kirill@phys.hawaii.edu}}
\affiliation{Department of Physics and Astronomy,
          University of Hawaii,\\ 2505 Correa Rd., Honolulu, Hawaii 96822}  
\author{Frank Petriello\thanks{frankjp@pha.jhu.edu}}
\affiliation{
Department of Physics, Johns Hopkins University, \\
3400 North Charles St., Baltimore, MD 21218
} 

\begin{abstract}
We present a calculation of the fully
differential cross section for  Higgs boson production 
in the gluon fusion channel
through next-to-next-to-leading 
order in perturbative QCD.  
We apply the method introduced 
in \cite{sector} to 
compute double real emission corrections. 
Our calculation permits {\it arbitrary} cuts on the final 
state in the reaction $hh \to H + X$. It can be easily extended 
to include decays
of the Higgs boson into observable final states. In this Letter, 
we discuss the most important features of the calculation, 
and present some examples of physical applications that illustrate the 
range of observables that can be studied using our result. 
We compute the NNLO rapidity distribution of the 
Higgs boson, and also calculate the NNLO rapidity distribution with a 
veto on jet activity.

\end{abstract}

\maketitle

\end{titlepage}

With Run II of the Tevatron producing data and the LHC set to begin operation in 2007, 
hadron colliders will soon play a major role in 
understanding the mechanism of electroweak symmetry breaking. 
Within the Standard Model, this mechanism is linked to the Higgs boson, 
a scalar particle whose non-zero vacuum 
expectation value gives rise to the masses of all elementary particles.
Finding the Higgs boson and analyzing 
its properties are therefore important parts of the high-energy physics 
program in the next decade. 

Hadron colliders, while offering substantial increases in energy 
over existing lepton machines, present several obstacles to performing 
precision physics studies.  
Non-perturbative QCD enters the calculation of cross sections 
through both parton distribution functions (pdfs) of hadrons 
in the initial state and properties of jets in the final state.  
Perturbatively, the large value of the QCD coupling constant and the 
enhanced sensitivity to the factorization and renormalization scales force 
calculations of many important processes to be extended to 
next-to-next-to-leading order (NNLO) in the QCD coupling constant.
This reduces the unphysical sensitivity of the result to the 
renormalization and factorization scales; also, 
the additional partons model the structures of both colliding hadrons and 
final-state jets more accurately.

There has been significant progress in the past several years in performing 
inclusive and semi-inclusive NNLO calculations 
\cite{Harlander:2002wh,Brein:2003wg,Harlander:2003ai,DY,Anastasiou:2003yy}.  
However, 
because of the cuts on the final states typical for the LHC and the 
Tevatron, such results are of limited use.  A fully differential, partonic 
Monte Carlo event generator is preferable.  It is then guaranteed that, 
to a given order in the perturbative 
expansion, there is full control over hard emissions and the normalization 
of a given observable.  However, this approach can not be applied if the 
observable is dominated by regions of the phase space where soft and collinear 
effects are enhanced, and the resummation of certain higher order 
corrections may be necessary.  
In such situations, shower Monte Carlo event generators are used;  
they correctly describe
the soft and collinear limits of the real emission. However, they can not  
reproduce the properties of hard radiation, and are unsuitable for 
precision studies.  
Ideally, we would combine perturbative results 
with shower Monte Carlos to gain 
the advantages of both approaches.  This has been achieved at NLO \cite{Frixione:2002ik}, 
but has not been extended to higher orders.  Constructing fully 
differential perturbative results is a first step toward this goal.      

Extending exclusive calculations to higher orders is not straightforward.
While at NLO this problem has been solved \cite{NLOdipole}, a similar 
solution at NNLO is not yet available, despite significant 
effort \cite{NNLOdipole}.  The primary obstacle is the double real radiation 
contribution to NNLO cross sections, which contains two additional 
partons in the final state.  The singularity structure resulting from 
these emissions is significantly more complex 
than at NLO, and prevents the use of NLO techniques.

We have recently suggested an alternative approach to the problem 
of real radiation at NNLO \cite{sector}, which allows differential 
results 
to be obtained in an efficient, automated fashion.  We have tested our idea on the 
realistic example of $e^+e^- \to 2~{\rm jets}$ at NNLO \cite{sector,Anastasiou:2004qd}.  
In this Letter we apply our method to calculate the fully 
differential cross section 
for Higgs boson production at hadron colliders through NNLO in QCD.  Since our 
result retains all kinematic information, it can be used to 
compute arbitrary differential distributions, or to construct a partonic 
event generator accurate to NNLO.

The dominant production mechanism for a light Higgs boson at hadron 
colliders is 
gluon fusion, $gg \to H$,  through a top quark loop. 
If the Higgs boson is sufficiently light, $m_h < 2m_t$, its coupling to 
gluons can be described by a point-like vertex. 
Within this approximation, the NNLO corrections to the inclusive cross-section 
for Higgs hadroproduction have been considered recently in 
\cite{Harlander:2002wh}, where 
a detailed description of the setup can be found.  
The two-loop virtual corrections required for Higgs production 
are identical for differential distributions and the total cross section.  
The one-loop corrections to single real-radiation processes 
($gg \to H+g$, $q \bar q \to H+g$, $qg \to H+g$) can be 
computed easily with established 
methods~\cite{Anastasiou:2003yy}.
The difficult contributions are the double real emission
corrections that first appear at NNLO.


We use the method first presented in \cite{sector} to calculate these components. 
We combine an expansion in plus distributions with sector decomposition \cite{decomp} to 
separate and extract their singularities.  This requires a parameterization of 
the phase space in which the integration region is the unit hypercube.  
In principle, any mapping that accomplishes this is acceptable.  In practice, 
finding a convenient parameterization that reduces the number of sector 
decompositions is important for the efficiency of the approach.

%
For the double real emission corrections to Higgs production at NNLO, 
we must parameterize a $2 \rightarrow 3$ particle phase space, with 
one massive final-state particle.  
We consider here 
$g(p_1) + g(p_2) \to H(h) + g(p_3) + g(p_4)$ as a prototypical partonic process.  
For a fixed energy of
the partonic collision $(p_1+p_2)^2 = s$, the partonic phase space is 
described by four independent variables.
We found it useful to employ several different parameterizations of 
the partonic phase space.  We present here explicit formulae only 
for the parameterization which is the 
most convenient choice for the majority of diagrams.
The scalar products $s_{34}=(p_3+p_4)^2$, $s_{ih}=(p_i-p_h)^2$, and $s_{2j}=(p_2-p_j)^2$, 
where $i=1,2$ and $j=3,4$, 
take on simple forms in this parameterization, while the $s_{1j}$ contain overlapping 
singularities, and require sector decomposition when appearing in the 
denominator.
In this parameterization, the $2 \rightarrow 3$ phase space becomes
\ba
&& {\rm d\Pi_R} = \int [dp_3][dp_4][dh] \delta^{(d)}(p_1+p_2 - h - p_3 - p_4)
\nonumber \\
&&= N \int \limits_{0}^{1} {\rm d}\lambda_1 {\rm d}\lambda_2 
{\rm d}\lambda_3 {\rm d}\lambda_4 \left[(1-\lambda_1)(1-\lambda_1 K_m/K_p)\right]^{-\epsilon} 
\nonumber \\ & & \times \left[\lambda_1\lambda_2(1-\lambda_2)\right]^{-\epsilon} 
\left[\lambda_3 (1-\lambda_3)\right]^{1-2\epsilon}
\left[\lambda_4(1-\lambda_4)\right]^{-\epsilon-1/2} \nonumber \\ & & \times
\left[K_p r/(1+u)^2\right]^{-1+\epsilon}\left[1-\frac{\lambda_1 K_m}{r(1+z)}\right],
\label{rapparam}
\ea
where $d = 4-2\epsilon$ is the space-time dimensionality, 
$[dq] = {\rm d}q^{d-1}/(2q_0)$, $z=m_h^2/s$,
$N = \Omega(d-2) \Omega(d-3) (1-z)^{3-4\epsilon}/2^{4+2\epsilon}$, 
$\Omega(d) = 2\pi^{d/2}/\Gamma(d/2)$,
and the independent scalar products are
\ba 
s_{1h} &=& -
\lambda_3 (1-z)\left[1-\lambda_1 r(1-rt)/(r+t)\right], \nonumber \\ 
s_{23} &=& -\lambda_2 \lambda_3 (1-z) \left[1+\lambda_1 (1-rt)/r/(r+t)\right], \nonumber \\ 
s_{34} &=& \lambda_1 \lambda_3 (1-\lambda_3) (1-z)^2 (1+u)^2/K_p/r, \nonumber \\ 
s_{13} &=& -\frac{(1-\lambda_3)(1-z)}
{K_p r \left[1+\lambda_1 (1-rt)/r/(r+t)\right]} \nonumber \\ & & \times 
\left[A_1+A_2+2(2\lambda_4-1)\sqrt{A_1A_2}\right].
\label{eqrap}
\ea 
We have introduced the notations
$u = (z+\lambda_3(1-z))/(1-\lambda_3(1-z)),~~r=\sqrt{u},~~~t=\sqrt{z},~~ 
K_{p,m} = (r \pm t)(1 \pm rt),~~
A_1 = \lambda_1(1-\lambda_2)(1+u)^2,~~
A_2 = \lambda_2(1-\lambda_1)r(K_p-\lambda_1 K_m)$.

With the expressions given above, it is in principle straightforward 
to apply the method explained in \cite{sector} to derive the 
finite, fully differential cross section for Higgs production at NNLO.  
However, some technical points warrant further discussion.
A brute force application of the algorithm in~\cite{sector} typically 
produces very large expressions.  It is therefore important to organize the 
calculation efficiently. 
\begin{itemize}
\item We found it important to identify all possible 
symmetries of the process, and utilize the fact that many terms in the 
matrix elements are identical under a simple rotation of momentum labels, in order to 
reduce expression sizes.

\item 
We found it useful to introduce a specific phase-space parameterization only for the denominators of the matrix elements.  
We keep the numerators written in terms of invariant masses.  We therefore 
divide the expressions into universal denominator structures (topologies), valid for any 
$2 \rightarrow 3$ real emission correction with one massive final-state particle, 
and process-dependent numerators.  This allows us to implement other processes in our code very simply.  

\item 
A few topologies contain denominators that depend quadratically upon the $\lambda_i$.  
Since the full matrix elements do not contain quadratic singularities, there must be 
numerator structures responsible for regulating this behavior.  We therefore cannot 
trivially keep the numerator written in terms of invariant masses for these terms.  It is usually 
simple to identify the required numerator structure.  
An example found in Higgs production 
is the topology
\be
\frac{{\cal A}(s_{ij})}{s_{23}^2 s_{14}^2} = 
\frac{(s_{13}s_{24}-s_{34})^2}{s_{23}^2s_{14}^2} \tilde {\cal A}(s_{ij}). 
\label{eqr}
\end{equation}
The left-hand side of Eq.(\ref{eqr}) naively has quadratic singularities, but, as shown,  we can identify the required scalar product in the numerator that regulates these.  
We can now write $\tilde {\cal A}(s_{ij})$ 
using invariant masses, and introduce 
a specific parameterization for the remainder.

\item As in previously studied examples \cite{sector}, the matrix elements 
contain ``line'' singularities
that arise when a singularity is mapped to an edge of phase space. 
These are removed using an additional variable change 
$\lambda_i \to \tilde \lambda_i$, as discussed in \cite{sector}.

\item When we convolute the partonic cross sections with the pdfs to form the 
hadronic cross section, the variable $z$ scales as 
$z \rightarrow m_{h}^2/(x_1x_2s_{had})$, where 
$s_{had}$ is the hadronic center-of-mass energy squared, and the $x_i$ are the 
fractions of the hadronic momenta carried into the hard scattering process.  The 
partonic cross sections are distributions in $z$, and are singular 
in the limit $z \rightarrow 1$.  Is is clear from Eq.(\ref{rapparam}) 
that the factor $(1-z)^{-4\epsilon}$ which regulates this limit has been extracted, 
and therefore singularities in $z$ and $\lambda_i$ can be treated identically.

\end{itemize}

After the extraction of singularities, we can combine the double real emission corrections 
with the remaining contributions to the hard scattering cross section.  
This expression is then integrated numerically together with the pdfs to form 
the hadronic cross section.  We perform the numerical integration with the 
version of VEGAS described in \cite{Hahn:2004fe}.  We use the MRST pdf sets, 
with {\sf mode}=1.  The numerical integration we perform is 6-dimensional; this 
includes the four independent partonic variables and the two $x_i$.  
The CPU 
times required for this calculation depend
strongly on both the kinematics and the requested precision.  For 
Tevatron kinematics, reaching 1\% precision on the inclusive cross section requires 
0.5 hours, while obtaining 1\% at the LHC requires about 1.5 hours.  When constraints 
are imposed on the final state, the run times increase.  


We have performed several checks on our calculation.  We cancel 
$1/\ep$ poles numerically, as is typically done using this method \cite{sector,Anastasiou:2004qd}.  
The singularities begin at $1/\ep^4$, and involve all the different 
contributions to the final result: two-loop virtual corrections, ultraviolet 
renormalization, virtual corrections to NLO processes, 
double real emission, and collinear subtractions for pdfs;
the cancellation of the singularities is therefore a strong 
check on the calculation. We find that the singularities 
cancel to a high precision for both the inclusive cross section and 
all distributions we have studied.  For example, the $\ep$-expansion for 
the NNLO inclusive cross section at the LHC is
\begin{eqnarray}
\sigma &=& (-0.05 \pm 2)\times 10^{-6}/\ep^4 + (-0.7 \pm 2)\times 10^{-3}/\ep^3
\nonumber \\ &+& (-0.3 \pm 2) \times 10^{-3}/\ep^2 + (-3 \pm 9)\times 10^{-3}/\ep 
\nonumber \\ &+& (44.9\pm 0.4).
\end{eqnarray}
These results are in picobarns, and are obtained using 
$m_h = 120$ GeV and $m_t = 175$ GeV; for this 
and other results in this paper, higher order QCD corrections are obtained using the 
effective Lagrangian valid for $m_h < 2m_t$ and then normalized to the exact tree-level cross 
section with its full $m_t$-dependence.  The renormalization and factorization scales have 
been set equal to the Higgs boson mass, $\mu_R=\mu_F=m_h$.
We have found a very good agreement between our result 
for the finite inclusive 
cross section for both Tevatron and LHC kinematics
and several calculations of this quantity 
available in the literature~\cite{Harlander:2002wh}.

\noindent
\begin{figure}[t]
\centerline{
\psfig{figure=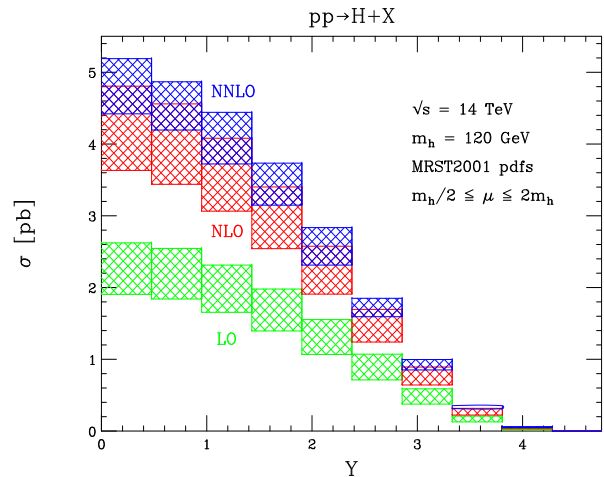,height=6.3cm,width=7.9cm,angle=90}}
\vspace{0.0cm}
\caption{Bin-integrated rapidity distribution for LHC kinematics.  The scale $\mu$ is varied 
between $m_h/2 \leq \mu \leq 2m_h$.  The LO, NLO, and NNLO distributions are shown.}
\label{rapplot}
\end{figure}
\vspace{-0.4cm}

We now present several distributions that illustrate the range of 
observables that can be studied using our computation.  
The numerical precision 
of all results shown is 1\%.  We first compute 
the bin-integrated rapidity distribution of the Higgs boson.  
We separate the entire rapidity range into twenty bins; because 
of the symmetry of the distribution under $Y \to -Y$, we only need to consider 
ten bins for $Y \ge 0$.  The resulting distribution is shown in 
Fig.\ref{rapplot}.
We have equated the renormalization and factorization scales, $\mu_R=\mu_F=\mu$, and have 
varied them in the range $m_h/2 \leq \mu \leq 2m_h$.  There are large corrections 
to the rapidity distribution; however, they arise primarily 
from the inclusive $K$-factor  $K=\sigma_{\rm NNLO}/\sigma_{\rm LO}$.
The rapidity dependence of the NNLO $K$-factor is 
small, and is insignificant if normalized to the NLO distribution.  
This is not unexpected; the production of the 
Higgs boson at the LHC is strongly dominated by the soft $z \to 1$ 
limit, which implies that the kinematics of the tree level process 
is not altered significantly by hard QCD emissions. 

\begin{figure}[t]
\centerline{
\psfig{figure=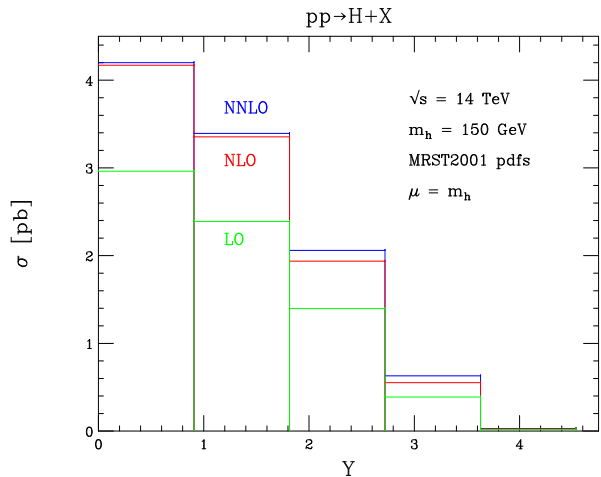,height=6.3cm,width=7.9cm,angle=90}}
\vspace{0.0cm}
\caption{Bin-integrated rapidity distribution for LHC kinematics, with a jet veto 
of $|p_{\perp}^{j}| < 40$ GeV.  We have set $\mu=m_h$, and have included the 
LO, NLO, and NNLO results}.  
\label{vtplot}
\end{figure}

In Fig.\ref{vtplot} 
we present the rapidity distribution with a veto on jet activity, 
as described in \cite{Catani:2001cr}.  We use the cone algorithm with $R = 0.4$, and 
use $p_{\perp}^{veto} = 40$ GeV; we therefore demand $|p_{\perp}^{j}| < 40$ GeV.  
For this result we set $m_h = 150$ GeV, as this cut 
is of particular importance for the $H \rightarrow W^{+}W^{-}$ decay channel.  We show 
the LO, NLO, and NNLO results; we note that the LO vetoed rapidity distribution is 
identical to the inclusive LO rapidity distribution, as there is no additional 
radiation at leading order in perturbation theory.  The 
$K$-factor  is much smaller for the vetoed cross 
section than for the inclusive cross section; this can be easily seen by comparing the results in Fig.\ref{rapplot} and Fig.\ref{vtplot}.  
One reason for this is that 
the average $p_{\perp}$ of the Higgs boson increases from NLO to NNLO; we find 
$<p_{\perp}^{\rm NLO}> = 37.5$ GeV, while 
$<p_{\perp}^{\rm NNLO}> = 44.6$ GeV.  Since the additional 
partons recoil against the Higgs, this is equivalent to an increase in the average 
jet $p_{\perp}^{j}$.  The increase of $p_{\perp}^{j}$ means that less of the cross section 
passes the veto, leading to a
larger reduction in the NNLO cross-section relative to the NLO one.

\noindent
\vspace{-0.7cm}

We have described a calculation of the fully differential cross section for 
Higgs hadro-production at NNLO.  We have utilized the method developed 
in \cite{sector} to extract and cancel double real radiation singularities, 
and have extended this approach to handle processes with initial-state 
collinear singularities.   Our result allows arbitrarily differential 
observables to be studied; as examples, we have presented results 
for the Higgs rapidity distributions at the LHC, both with and 
without a veto on jet activity.  This approach can be easily extended 
to include decays of the Higgs, or can be applied to other 
$2 \rightarrow 1$ processes of phenomenological interest.  
We will
provide  a detailed description of our calculation  and 
present a public version of our numerical code in a 
forthcoming publication.


We thank T. Hahn for helpful communications regarding his 
numerical integration package CUBA.
This work was started when the authors were visiting the
Kavli Institute for Theoretical 
Physics, UC California, Santa Barbara. C.A. would like 
to thank ETH, CERN, Saclay, Fermilab and the Johns Hopkins University 
for their hospitality.
This research was supported by the US Department of Energy  under contracts
DE-AC02-76SF0515, DE-FG03-94ER-40833 and 
the Outstanding Junior Investigator Award DE-FG03-94ER-40833, and
by the National Science 
Foundation  under contracts P420D3620414350, P420D3620434350 
and partially under Grant No. PHY99-07949.



\begin{thebibliography}{29}

\bibitem{sector}
C.~Anastasiou, 
K.~Melnikov and F.~Petriello,
Phys.\ Rev.\ D {\bf 69}, 076010 (2004).


\bibitem{Harlander:2002wh}
R.~V.~Harlander and W.~B.~Kilgore,
Phys.\ Rev.\ Lett.\  {\bf 88}, 201801 (2002);
%
C.~Anastasiou and K.~Melnikov, 
Nucl.\ Phys.  {\bf B646}, 220 (2002);
%
V.~Ravindran, 
 J.~Smith and W.~L.~van Neerven,
Nucl.\ Phys. {\bf B665}, 325 (2003).

\bibitem{Brein:2003wg}
O.~Brein {\it et al.},
Phys.\ Lett.  {\bf B579}, 149 (2004).

\bibitem{Harlander:2003ai}
R.~V.~Harlander, 
and W.~B.~Kilgore,
Phys.\ Rev. {\bf D68}, 013001 (2003).

\bibitem{DY}
R.~Hamberg {\it et al.},
Nucl.\ Phys.\ B {\bf 359}, 343 (1991)
[Erratum-ibid.\ B {\bf 644}, 403 (2002)].

\bibitem{Anastasiou:2003yy}
C.~Anastasiou,
L.~J.~Dixon, K.~Melnikov and F.~Petriello,
Phys.\ Rev.\ Lett.\  {\bf 91}, 182002 (2003);
%
Phys.\ Rev.\ D {\bf 69}, 094008 (2004).


\bibitem{Frixione:2002ik}
S.~Frixione {\it et al.},
JHEP {\bf 0206}, 029 (2002).

\bibitem{NLOdipole}
R.~K. Ellis {\it et al.}, Nucl. Phys. {\bf B178}, 421 (1981);
W.~T. Giele and E.~W.~N. Glover, Phys. Rev. {\bf D46}, 1980 (1992);
K.~Fabricius {\it et al.}, Phys. Lett. {\bf B97}, 
431 (1980); F.~Gutbrod, G.~Kramer and G.~Schierholz, Z. Phys. {\bf C21}, 
235 (1984);
Z.~Kunszt and D.~E.~Soper,
Phys.\ Rev.\ D {\bf 46}, 192 (1992);
S.~Frixione, Z.~Kunszt and A.~Signer,
Nucl.\ Phys.\ B {\bf 467}, 399 (1996);
S.~Catani and M.~H.~Seymour,
Nucl.\ Phys.\ B {\bf 485}, 291 (1997)
[Erratum-ibid.\ B {\bf 510}, 503 (1997)].

\bibitem{NNLOdipole}
J.~M.~Campbell and E.~W.~N.~Glover,
Nucl.\ Phys.\ B {\bf 527}, 264 (1998);
D.~A.~Kosower,
Phys.\ Rev.\ D {\bf 67}, 116003 (2003);
S.~Weinzierl,
JHEP {\bf 0303}, 062 (2003);
A.~Gehrmann-De Ridder {\it et al.},
hep-ph/0403057;
W.~B.~Kilgore,
hep-ph/0403128;
A.~Gehrmann-De Ridder, T.~Gehrmann and E.~W.~N.~Glover,
hep-ph/0407023.


\bibitem{Anastasiou:2004qd}
C.~Anastasiou,
K.~Melnikov and F.~Petriello,
Phys.\ Rev.\ Lett.\  {\bf 93}, 032002 (2004).

\bibitem{decomp}
T.~Binoth and G.~Heinrich,
Nucl.\ Phys.\ B {\bf 585}, 741 (2000); for earlier work see 
K.~Hepp,
Commun.\ Math.\ Phys.\  {\bf 2}, 301 (1966);
M.~Roth and A.~Denner,
Nucl.\ Phys.\ B {\bf 479}, 495 (1996).

\bibitem{Hahn:2004fe}
T.~Hahn,
hep-ph/0404043.

\bibitem{Catani:2001cr}
S.~Catani {\it et al.},
JHEP {\bf 0201}, 015 (2002).


\end{thebibliography}
\end{document}